\documentclass[twocolumn,prl,showpacs]{revtex4}
\begin{document}

\topmargin-1.0cm

\title
{Real-time electron dynamics with exact-exchange time-dependent density-functional theory}
\author {H. O. Wijewardane}
\altaffiliation[Present address: ]{Department of Physics and Astronomy,
Hunter College and CUNY, New York, New York 10021}

\author{C. A. Ullrich}

\affiliation
{Department of Physics and Astronomy, University of Missouri, Columbia, Missouri 65211}
\date{\today}

\begin{abstract}
The exact exchange potential in time-dependent density-functional theory  is defined as an orbital functional
through the time-dependent optimized effective potential (TDOEP) method.
We numerically solve the TDOEP integral equation for the real-time nonlinear intersubband
electron dynamics in a semiconductor quantum well with two occupied subbands. By comparison with adiabatic approximations,
it is found that memory effects in the exact exchange potential become significant when the electron dynamics takes place
in the vicinity of intersubband resonances.
\end{abstract}
\pacs{31.15.Ew, 71.15.Mb, 71.45.Gm}
\maketitle

Time-dependent density-functional theory (TDDFT) \cite{Runge1994,tddft} is an increasingly popular approach to
the electron dynamics in atoms, molecules and solids. Much of TDDFT's success is due to
its ability to describe electronic excitations in large molecules \cite{Casida1995,Elliott2007} using
simple time-dependent exchange-correlation (XC) functionals such as the adiabatic local-density approximation
(ALDA). However,  neither the ALDA nor any adiabatic gradient-corrected XC functional captures multiple or charge-transfer
excitations \cite{Maitra2004, Hieringer2006} or excitons \cite{Reining2002}.
These difficulties are not caused by intrinsic deficiencies of TDDFT, but indicate the need for better
functionals: for multiple and charge-transfer excitations one must abandon the adiabatic approximation, and
excitons in solids require long-range XC kernels with a $1/q^2$ behavior. Moreover, for
strong-field multiple ionization it was suggested that the XC potential should change discontinuously
with the number of electrons $N$ \cite{Lein2005}.

Several promising ideas for new time-dependent XC functionals have been explored, such as current-TDDFT
\cite{Vignale1997,Wijewardane2005} and many-body Green's function techniques \cite{Reining2002}.
Each of these approaches has its advantages and drawbacks:
current-TDDFT is well-suited to describe polarizability and collective excitations in extended
systems \cite{Ullrich2001,Faassen2002}, but introduces spurious dissipation in finite systems
\cite{Ullrich2004}. The many-body techniques of Refs. \cite{Reining2002} give excellent optical
spectra in insulators, but cannot be easily extended into the nonlinear or the real-time domain.

In static DFT, orbital-dependent functionals have long been recognized as powerful and versatile
\cite{Goerling2005}. If the XC energy is given in terms of the orbitals (e.g., the exact
exchange energy), the associated local XC potential is constructed with the optimized effective potential (OEP) method \cite{Talman1976}
or by the simplified but nearly as accurate KLI scheme \cite{Krieger1992}. This leads to XC potentials
that are self-interaction free, have the correct $-1/r$  asymptotics for
finite systems, exhibit discontinuities upon change of $N$, and generally produce high-quality
orbitals, eigenvalues and band structures \cite{Goerling2005}.

This paper deals with the time-dependent optimized effective potential (TDOEP) \cite{Ullrich1995,RvL1996}, which
generalizes the static OEP method, carrying over the desirable properties mentioned above and introducing new features
that are unique to the dynamical case.
The TDOEP integral equation for the local XC potential $V_{\rm {xc}\sigma}$ is given by \cite{Ullrich1995}
\begin{eqnarray}\label{tdoep}
0 &=&
i\sum_{j=1}^{N{\sigma}} \int_{-{\infty}}^t{d}t'\int{d}^3r'[V_{\rm {xc}{\sigma}}({\bf r}',t')-u_{{\rm xc}j{\sigma}}
({\bf r}',t')]\nonumber\\
&\times& \sum_{k=1}^{\infty}
{\phi}_{k{\sigma}}({\bf r}',t')
{\phi}_{k{\sigma}}^*({\bf r},t)
{\phi}_{j{\sigma}}^*({\bf r}',t'){\phi}_{j{\sigma}}({\bf r},t)+c.c.
\end{eqnarray}
The $\phi_{j\sigma}$ are time-dependent Kohn-Sham orbitals,
\begin{equation} \label{uxcj}
u_{{\rm xc}j\sigma}({\bf r},t)=\frac{1}{{\phi}_{j{\sigma}}^*({\bf r},t)}\frac{{\delta}A_{\rm xc}}
{\delta{\phi}_{j{\sigma}}({\bf r},t)}\:,
\end{equation}
and $A_{\rm xc}$ is an orbital-dependent XC action functional (which is rigorously defined on a Keldysh contour
\cite{RvL1996}). Here and in the following, we formally include a spin index $\sigma$, but consider only nonmagnetic systems.

Linearization of the TDOEP approach leads to a frequency-dependent XC kernel \cite{Goerling1998}, which was successfully
applied to optical absorption spectra of insulators \cite{Kim2002} and dynamic polarizabilities of atoms
\cite{Hirata2006}. In these applications, the frequency dependence of the linearized XC potential was found to play only a minor role.

Until now there have been no applications of the full, real-time TDOEP scheme, only of the TDKLI approximation
\cite{tddft,Ullrich1995}.
A previous unsuccessful attempt at solving Eq. (\ref{tdoep}) was plagued by
numerical instabilities  \cite{Mundt2006}. In this paper, we present an algorithm for stable numerical solutions of
the exact-exchange TDOEP integral equation (\ref{tdoep}), and apply it to the nonlinear electron dynamics in quantum wells.

Our goal is to explore the significance of memory effects in TDOEP.
Memory-dependent XC potentials in current-TDDFT cause elastic and dissipative effects in the electron dynamics
\cite{Wijewardane2005}. In particular, the adiabatic approximation was shown to break down in the limit of
large, rapid deformations \cite{UllrichTokatly2006}. Here, we study the so far unresolved question of the importance of
memory in exact-exchange TDDFT in different dynamical regimes. To isolate the effects of memory in a clear-cut way,
we will compare the full TDOEP with an adiabatic approximation (AOEP).

{\em Adiabatic TDOEP scheme.}
Under the implicit assumption that a system evolves so slowly that it always remains close to the ground state of
a given time-dependent potential, adiabatic XC functionals have no memory and only
depend on the instantaneous density or orbitals. While this is straightforward in ALDA and TDKLI, the AOEP
is more complicated since the static OEP \cite{Talman1976,Krieger1992} also depends on the energy eigenvalues,
whose meaning in the dynamical regime is not obvious.

We define the AOEP  at time $t$ as that static OEP potential whose associated ground-state density equals the instantaneous
density $n(t)$. This requires two steps: First, find that static Kohn-Sham potential $V_{\rm KS}^t$
(e.g., by iteration \cite{RvL1994})
which produces $n(t)$ as its ground-state density, and obtain the associated complete set of orbitals $\phi_{j\sigma}^t({\bf r})$
and eigenvalues $\epsilon_{j\sigma}^t$. Next, take these $\phi_{j\sigma}^t({\bf r})$ and
$\epsilon_{j\sigma}^t$ and plug them into the
static OEP integral equation \cite{Talman1976,Krieger1992}. This yields the AOEP XC potential at time $t$.

{\em TDOEP for quantum wells.} We consider conduction electrons in $n$-doped semiconductor quantum wells in effective-mass
approximation \cite{tddft,Wijewardane2005,Ullrich2001}, confined along $z$.
The ground-state envelope function for the $j$th subband $\phi^0_{j\sigma} (z)$ follows from a one-dimensional
Kohn-Sham equation, with density $ n(z) =  \sum_{j}^{\rm occ} |\phi_{j\sigma}^0 (z)|^2 (\epsilon_F -\epsilon_{j\sigma})/\pi$
and subband and Fermi energy levels $\epsilon_{j\sigma}$ and $\epsilon_F$. To describe intersubband dynamics preserving the
translational symmetry in the quantum well plane (ignoring disorder and phonons), we propagate the subband envelope functions using
the time-dependent Kohn-Sham equation
\begin{eqnarray} \label{tdks}
i \frac{\partial}{\partial t} \phi_{j\sigma} (z,t) &=& \Big[
-  \frac{1}{2} \frac{\partial^2}{\partial z^2} + V_{\rm dr}(z,t)+ V_{\rm conf}(z)  \nonumber\\
&&{}+  V_{\rm H}(z,t) +V_{\rm xc \sigma}(z,t)\Big] \phi_{j\sigma} (z,t) \:,
\end{eqnarray}
with initial condition $\phi_{j\sigma}(z,t_0) = \phi^0_{j\sigma}(z)$.
Here, $V_{\rm dr}$ is a time-dependent driving field (see below), $V_{\rm conf}(z)$ and $V_{\rm H}$ are the confining square-well
and the Hartree potential.

Assuming that the system is in its ground state for $t<t_0$, one obtains the following TDOEP equation:
\begin{eqnarray}\label{tdoepwell}
0 &=&
i\sum_{j=1}^{N{\sigma}}(k_F^{j{\sigma}})^2 \int_{t_0}^t{d}t'\int dz'[V_{\rm {x}{\sigma}}(z',t')-u_{{\rm x}j{\sigma}}
(z',t')]\nonumber\\
&&
\times \sum_{k\ne j}
{\phi}_{k{\sigma}}(z',t'){\phi}_{k{\sigma}}^*(z,t) {\phi}_{j{\sigma}}^*(z',t'){\phi}_{j{\sigma}}(z,t)+c.c. \nonumber\\
&+&
\sum_{j=1}^{N{\sigma}}(k_F^{j{\sigma}})^2\int{d}z'[V^0_{{\rm x}{\sigma}}(z')-u^0_{{\rm x}j{\sigma}}(z')]
\nonumber\\
&&
\times \sum_{k{\neq}j} {\phi}_{k{\sigma}}^*(z,t) \frac{{\phi}_{j{\sigma}}^{0*}(z'){\phi}^0_{k{\sigma}}(z')}
{{\epsilon}_{j{\sigma}}-{\epsilon}_{k{\sigma}}} \: {\phi}_{j{\sigma}}(z,t)
+c.c. \:,
\end{eqnarray}
where $(k_F^{j{\sigma}})^2=2({\epsilon}_F-{\epsilon}_{j\sigma})$, $u_{{\rm x}j{\sigma}}$ for
quantum wells is obtained following Proetto {\em et al.} \cite{Proetto2003}, and $V_{\rm x\sigma}^0$, $u^0_{{\rm x}j\sigma}$ and
$\phi^0_{j\sigma}$ follow from a static OEP calculation. The second term in Eq. (\ref{tdoepwell}) comes from
${\phi}_{j{\sigma}}(z,t)= {\phi}_{j{\sigma}}^0(z)e^{-i\epsilon_jt}$ for $t < t_0$.

{\em Numerical algorithm.} Our numerical TDOEP approach  uses a uniform spatial grid along $z$ and a time discretization in uniform steps
${\Delta}t$, from $t_0$ up until some final time $T$. Let us first consider two separate problems:
(i) Assuming that a $V'_{\rm {x}{\sigma}}(z,t)$ is explicitly given for $t_0 \le t \le T$, Eq. (\ref{tdks}) is easily propagated with the
standard Crank-Nicholson and predictor-corrector schemes \cite{tddft}. This yields the orbitals $\phi'_{j\sigma}(z,t)$.
(ii) In turn, assuming some $\phi''_{j\sigma}(z,t)$'s to be given for $t_0 \le t \le T$, Eq. (\ref{tdoepwell}) can be
solved by discretizing the spatial and time integrals (e.g. with the trapezoidal rule),
which leads to a linear equation determining $V''_{\rm {x}{\sigma}}(z,t)$.

The full TDOEP scheme requires the simultaneous solution of Eqs. (\ref{tdks}) and (\ref{tdoepwell}) over the interval $[t_0,T]$.
We achieve this using a straightforward iterative loop, taking the orbitals calculated in step (i) as input to step (ii),
and then feeding the resulting XC potential back as input to step (i). Selfconsistency is reached if
$\phi''_{j\sigma} = \phi'_{j\sigma}$ and $V''_{\rm {x}{\sigma}}=V'_{\rm {x}{\sigma}}$. The loop is initialized with
the TDKLI approximation for $V'_{\rm {x}{\sigma}}$ in the first iteration step.

Let us define $G_l=\int_{t_0}^T{d}t|d_{l}(t)-d_{l-1}(t)| / \int_{t_0}^T{d}t|d_{l}(t)|$ to monitor convergence, where
$d_l(t)=\int zn_l(z,t)dz$ is the dipole moment at the $l$th iteration. Our algorithm is stable and rapidly convergent,
as we will show below.

\begin{figure}
\unitlength1cm
\begin{picture}(5.0,7.5)
\put(-8.5,-13.3){\makebox(5.0,7.5){\includegraphics{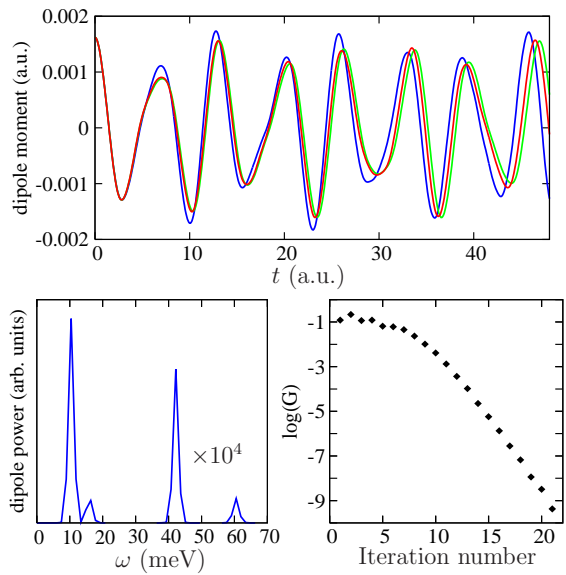}}}
\end{picture}
\caption{\label{figure1}(Color online) Top: dipole moment of free charge-density oscillations in a
quantum well with two occupied subbands [Blue, red, green (dark, medium, light gray): TDOEP, AOEP, TDKLI].
Bottom left: TDOEP dipole power spectrum. Bottom right: convergence index $G$ (see text).}
\end{figure}

{\em Results and discussion.} We consider  a 40-nm square GaAs/Al$_{0.3}$Ga$_{0.7}$As quantum well
with conduction band effective mass $m^* = 0.067m$ and charge $e^* = e/\sqrt{13}$, and with an electronic
density $N_s=2.2{\times}10^{11}{\rm cm}^{-2}$ such that the two lowest subbands $(j=1,2)$ are occupied.

We begin with the case of free charge-density oscillations. The initial state is calculated in the presence of
a 0.01mV/nm static electric field (``tilted'' quantum well). At $t_0=0$, the field is abruptly switched
off, which puts the electrons in an excited state and triggers collective charge-density oscillations.
Fig. \ref {figure1} shows that the convergence index $G$ of our numerical algorithm drops down
to $10^{-9}$ after only 20 iterations. Convergence was similar in all our TDOEP calculations.
We found that the zero-force theorem was always satisfied to within the limits of numerical resolution,
even for TDKLI (despite recent reports to the contrary in  metal clusters \cite{Mundt2007}). The
strong quantum well confinement seems to help enforce the zero-force theorem in TDKLI.

Fig. \ref{figure1} shows the dipole moment $d(t)$ obtained with TDOEP, AOEP and TDKLI, as well as the TDOEP dipole power spectrum
(the others are very similar). The dynamics is dominated by the $1\!\to\! 2$ and $2\!\to \! 3$ intersubband plasmons;
the former dominates since 90\% of the electrons sit in the first subband at the given $N_s$.
Higher plasmons ($1 \!\to \!4$, $2 \!\to  \!5$) are orders of magnitude weaker.
The differences between the methods are minor: the TDOEP charge-density oscillations are slightly faster, TDKLI is slowest,
and AOEP in between, but much closer to TDKLI. The first plasmon frequency $\omega_{12}$ is 10.4 meV in TDOEP and 10.2 meV in
AOEP and TDKLI; $\omega_{23}=16.0$ meV in all three approaches, and $\omega_{14}$ and $\omega_{25}$ are at 42 and 61 meV, respectively.

We find that the exact-exchange TDOEP does not cause any dissipation, similar to the high-frequency limit of
current-TDDFT \cite{Vignale1997,Wijewardane2005}; the nonadiabatic XC contribution is thus purely elastic, i.e. phase shifted by $\pi$ with
respect to the adiabatic part \cite{UllrichTokatly2006}. This is consistent with the observed behavior of the TDOEP versus AOEP, where
the memory leads to a small frequency renormalization (blueshift) of the dominant $\omega_{21}$ plasmon.

Nonadiabatic effects play a pronounced role at high frequencies, when the system rapidly undergoes
large deformations. The crossover from the low- to the high-frequency region was found to occur around the
average plasma frequency of the system \cite{UllrichTokatly2006}. To explore different dynamic regimes, we now consider
charge-density oscillations driven by $V_{\rm dr}(z,t) = e{\cal E} z f(t)\sin(\omega t)$, with
electric field amplitude $\cal E$, intensity $I\sim {\cal E}^2$, and frequency $\omega$. The envelope $f(t)$
is switched on at initial time $t_0=0$ over a 1-cycle linear ramp and then kept constant throughout the propagation.

\begin{figure}
\unitlength1cm
\begin{picture}(5.0,6.5)
\put(-8.5,-13.4){\makebox(5.0,6.5){\includegraphics{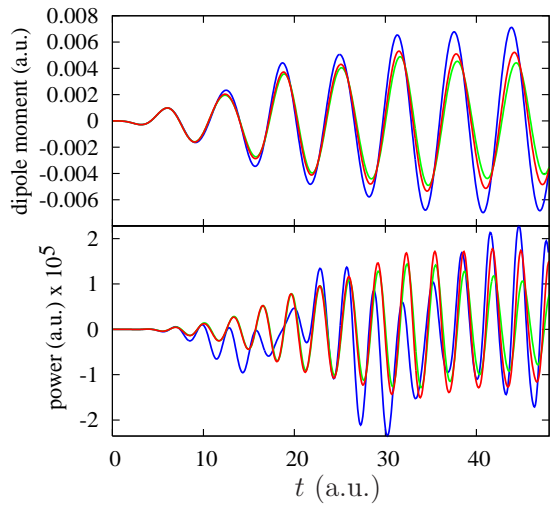}}}
\end{picture}
\caption{\label{figure2} (Color online)
Top: Dipole oscillations driven by an external field switched on at $t=0$, with frequency $\omega=11.2$ meV and intensity $I=10 \: \rm W/cm^2$.
Bottom: XC power [Eq. (\ref{power})]. Blue, red, green (dark, medium, light gray): TDOEP, AOEP, TDKLI.
}
\end{figure}

\begin{figure}
\unitlength1cm
\begin{picture}(5.0,6.5)
\put(-8.5,-13.4){\makebox(5.0,6.5){\includegraphics{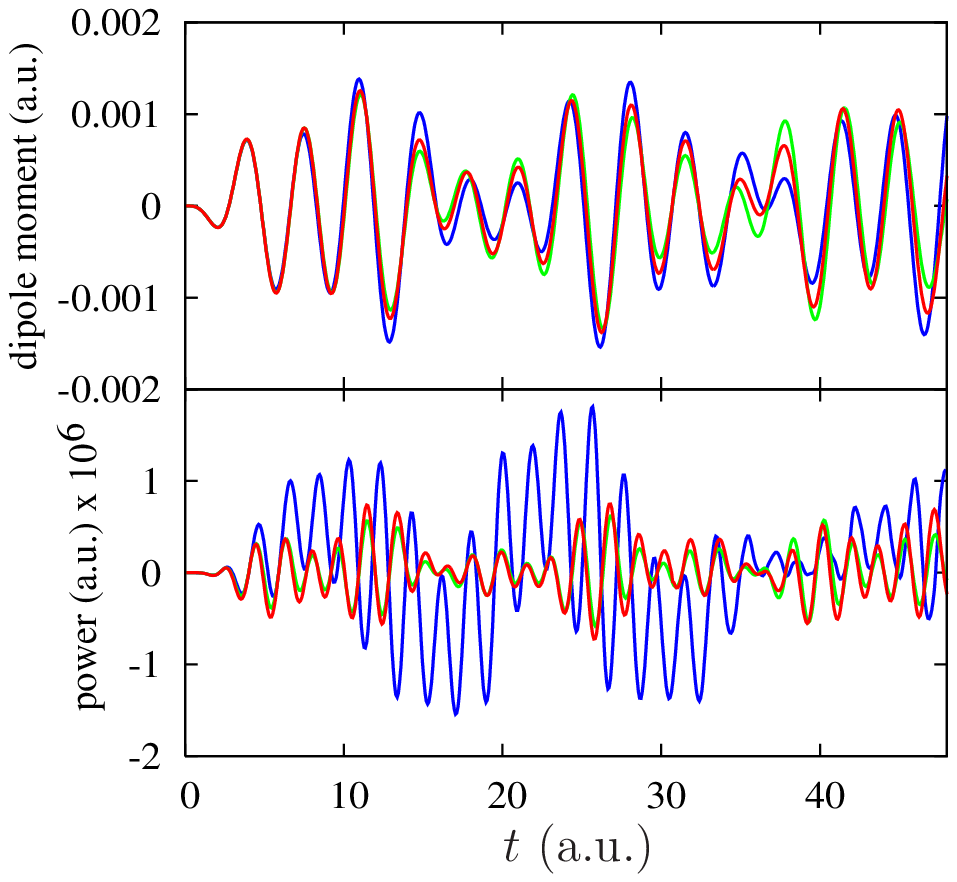}}}
\end{picture}
\caption{\label{figure3}
Same as Fig. \ref{figure2}, with $\omega=20$ meV and $I=20 \: \rm W/cm^2$.}
\end{figure}

\begin{figure}
\unitlength1cm
\begin{picture}(5.0,6.5)
\put(-8.5,-13.4){\makebox(5.0,6.5){\includegraphics{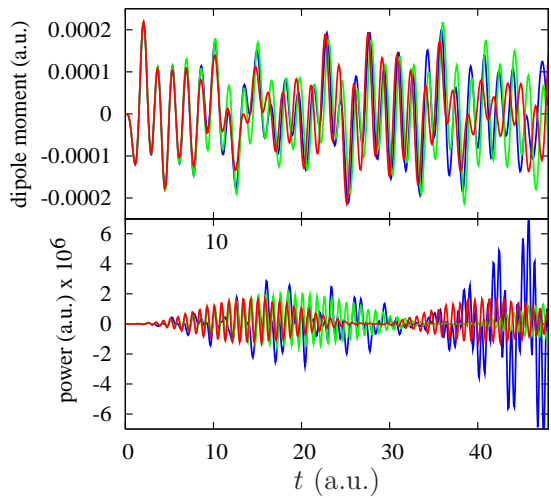}}}
\end{picture}
\caption{\label{figure4}
Same as Fig. \ref{figure2}, with $\omega=40$ meV and $I=40 \: \rm W/cm^2$.}
\end{figure}

\begin{figure}
\unitlength1cm
\begin{picture}(5.0,6.5)
\put(-8.5,-13.4){\makebox(5.0,6.5){\includegraphics{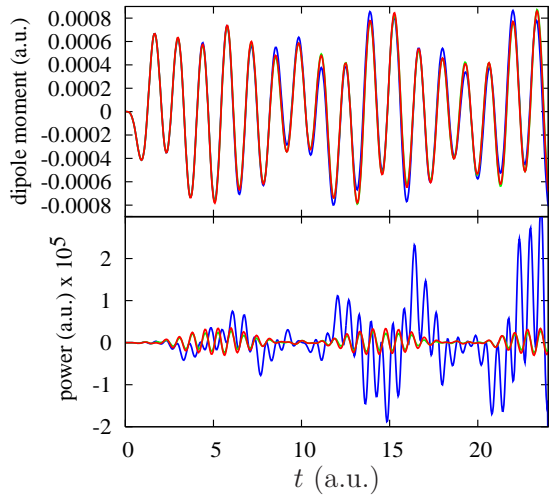}}}
\end{picture}
\caption{\label{figure5}
Same as Fig. \ref{figure2}, with $\omega=50$ meV and $I=1 \: \rm kW/cm^2$.}
\end{figure}

Figs. \ref{figure2}, \ref{figure3}, \ref{figure4}, and \ref{figure5} show $d(t)$ for $\omega= 11.2$, 20, 40,
and 50 meV, probing the dynamics close to $\omega_{12}$, $\omega_{23}$ and $\omega_{14}$, and between  $\omega_{14}$
and $\omega_{25}$. The intensities are 10, 20, 40, and 1000 $\rm W/cm^2$ respectively. As expected, the dipole
response is largest at $11.2$ meV since we are close to the dominating plasmon. At 50 meV
we need a much larger intensity since we are quite far away from any resonance.

Again, we find that the three methods give comparable results for $d(t)$. Figs. \ref{figure2}
and \ref{figure3} show that AOEP falls in between TDOEP and TDKLI, but remains closer to TDKLI. For $\omega=40$
meV there are more pronounced differences, probably due to the crosstalk of several
plasmon resonances slightly off tune. Interestingly, at $\omega=50$ meV, the results for $d(t)$ are very close.
We found a similar behavior at low frequencies well below $\omega_{12}$. This suggests that at the intermediate
frequencies we are studying here (not too far from the lowest intersubband plasmons), memory effects play a significant
role only in the vicinity of resonances.

For a more detailed analysis, especially of the phase shifts of $V_{\rm x\sigma}$, it is useful to consider the power
associated with the dynamic XC force:
\begin{equation} \label{power}
P(t)=\int{d}zj(z,t)\nabla_z\left[V_{{\rm x\sigma}}(z,t)-V^0_{{\rm x\sigma}}(z) \right],
\end{equation}
where $ j(z,t) = \sum_{k}^{\rm occ} \Im[\phi_{k\sigma}^*(z,t)\nabla_z \phi_{k\sigma}(z,t)]
(\epsilon_F -\epsilon_{k\sigma})/\pi$ is the current density. Since there is no dissipation, $P(t)$
is zero on average, but fluctuates more rapidly than the charge-density oscillations, at least twice as fast
as $d(t)$.

The bottom panels of Figs. \ref{figure2}-\ref{figure5} show $P(t)$ for $\omega=11.2$, 20, 40, and
50 meV. The differences between TDOEP versus AOEP and TDKLI are now much more apparent, in particular for higher
frequencies. At 20 meV one sees that $V_{\rm x\sigma}$ picks up a significant phase shift compared
to AOEP and TDKLI, which themselves are completely in sync. This is a clear indication of the elasticity induced
by the memory. The effect becomes even more pronounced at 40 and 50 meV.

We also find that $P(t)$ exhibits characteristic beating patterns (in particular at 40 meV), which are due to the detuning between the
driving field and the plasmon resonances. These patterns look similar in AOEP and TDKLI, but exhibit
marked additional structures in TDOEP.
Remarkably, as seen most clearly at 50 meV, these strong and rapid fluctuations of $V_{\rm x\sigma}$  in
TDOEP leave hardly any imprint on $d(t)$.

{\em Conclusion.} We have analyzed the electron dynamics in quantum wells with exact-exchange TDOEP, using a stable and rapidly convergent
numerical scheme. Away from resonances, the AOEP and TDKLI closely agree with each other
(like ground-state OEP and KLI \cite{Krieger1992}) and are good approximations to the full TDOEP.
Memory effects become more significant in the vicinity of intersubband resonances, resulting
in additional elastic contributions to the dynamics. Exact-exchange TDOEP has no memory at all in systems with only one occupied level
\cite{Krieger1992}, and our quantum well has only a small population of the second subband. We expect nonadiabatic effects to become
more important in systems with a larger relative occupancy of upper levels, and
in frequency regimes extending further beyond the lowest excitations.

Our analysis of the dynamic XC force shows that the  exact-exchange $V_{\rm x\sigma}$ has much richer temporal features
than the adiabatic approximations, but this has relatively little impact on the electron
dynamics itself. This suggests that for intrinsically nonadiabatic effects such as multiple excitations
\cite{Maitra2004} and dissipation \cite{Wijewardane2005} one needs to go beyond exact exchange.

\acknowledgments
This work was supported by Research Corporation and by NSF Grant No. DMR-0553485. We thank Paul de Boeij for valuable comments.


\begin{thebibliography}{100}

\bibitem{Runge1994}
E. Runge and E. K. U. Gross, Phys. Rev. Lett. {\bf 52}, 997 (1984).

\bibitem{tddft}
{\em Time-dependent density functional theory}, edited by M. A. L. Marques {\em et al.},
Lecture Notes in Physics {\bf 706} (Springer,  Berlin, 2006).

\bibitem{Casida1995}
M. E. Casida, in {\em Recent advances in density functional methods I}, ed. by D. E. Chong
(World Scientific, Singapore, 1995), p. 155.

\bibitem{Elliott2007}
P. Elliott {\em et al.},
arXiv:cond-mat/0703590 (2007).

\bibitem{Maitra2004}
N. T. Maitra {\em et al.},
J. Chem. Phys. {\bf 120}, 5932 (2004).

\bibitem{Hieringer2006}
W. Hieringer and A. Goerling, Chem. Phys. Lett. {\bf 419}, 557 (2006).

\bibitem{Reining2002}
L. Reining {\em et al.},
Phys. Rev. Lett. {\bf 88}, 066404 (2002);
F. Sottile {\em et al.}, Phys. Rev. Lett. {\bf 91}, 056402 (2003).

\bibitem{Lein2005}
M. Lein and S. K\"ummel, Phys. Rev. Lett. {\bf 94}, 143003 (2005); M. Mundt and S. K\"ummel,
Phys. Rev. Lett. {\bf 95}, 203004 (2005).

\bibitem{Vignale1997}
G. Vignale {\em et al.}
Phys. Rev. Lett. {\bf 79}, 4878 (1997).

\bibitem{Wijewardane2005}
H. O. Wijewardane and C. A. Ullrich, Phys. Rev. Lett. {\bf 95}, 086401 (2005).

\bibitem{Ullrich2001}
C. A. Ullrich and G. Vignale, Phys. Rev. Lett. {\bf 87}, 037402 (2001).

\bibitem{Faassen2002}
M. van Faassen {\em et al.},
Phys. Rev. Lett. {\bf 88}, 186401 (2002).

\bibitem{Ullrich2004}
C. A. Ullrich and K. Burke, J. Chem. Phys. {\bf 121}, 28 (2004);
C. A. Ullrich, J. Chem. Phys. {\bf 125}, 234108 (2006).

\bibitem{Goerling2005}
A. Goerling, J. Chem. Phys. {\bf 123}, 062203 (2005).

\bibitem{Talman1976}
J. D. Talman and W. F. Shadwick, Phys. Rev. A {\bf 14}, 36 (1976).

\bibitem{Krieger1992}
J. B. Krieger {\em et al.},
Phys. Rev. A {\bf 45}, 101 (1992).


\bibitem{Ullrich1995}
C. A. Ullrich {\em et al.},
Phys. Rev. Lett. {\bf 74}, 872 (1995).

\bibitem{RvL1996}
R. van Leeuwen, Phys. Rev. Lett. {\bf 76}, 3610 (1996) and Phys. Rev. Lett. {\bf 80}, 1280 (1998).

\bibitem{Goerling1998}
A. Goerling, Phys. Rev. A {\bf 57}, 3433 (1998).

\bibitem{Kim2002}
Y. -H. Kim and A. G\"orling, Phys. Rev. Lett. {\bf 89}, 096402 (2002).

\bibitem{Hirata2006}
Y. Shigeta {\em et al.}
Phys. Rev. A {\bf 73}, 010502(R) (2006).

\bibitem{Mundt2006}
M. Mundt and S. K\"ummel, Phys. Rev. A {\bf 74}, 022511 (2006).

\bibitem{UllrichTokatly2006}
C. A. Ullrich and I. V. Tokatly, Phys. Rev. B {\bf 73}, 235102 (2006).

\bibitem{RvL1994}
R. van Leeuwen and E. J. Baerends, Phys. Rev. A {\bf 49}, 2421 (1994).

\bibitem{Proetto2003}
F. A. Reboredo and C. R. Proetto, Phys. Rev. B {\bf 67}, 115325 (2003); S. Rigamonti and C. R. Proetto,
Phys. Rev. Lett. {\bf 98}, 066806 (2007).

\bibitem{Mundt2007}
M. Mundt {\em et al.},
Phys. Rev. A {\bf 75}, 050501(R) (2007).


\end{thebibliography}
\end{document}